\documentclass{PoS}

\title{Full jet reconstruction in 200 GeV p+p, d+Au and Au+Au collisions by STAR}

\ShortTitle{Jets in 200 GeV p+p, d+Au and Au+Au collisions}

\author{\speaker{Jan KAPITAN} for the STAR Collaboration \\
        Nuclear Physics Institute, Academy of Sciences of the Czech Republic\\
        E-mail: \email{kapitan@rcf.rhic.bnl.gov}}

\abstract{

Measurements of inclusive hadron suppression and di-hadron azimuthal correlations have provided important insights into jet quenching in hot QCD matter. However, they do not provide access to the energy of the hard scattering and are limited in their sensitivity since they can be affected by biases toward hard fragmentation and small energy loss. 

Full jet reconstruction in heavy-ion collisions enables a complete study of the modification of jet structure due to energy loss, but is challenging due to the high multiplicity environment.
Study of jet production and properties in d+Au and p+p collisions provides important baseline measurement for jet studies in heavy-ion collisions.

We report measurements of fully reconstructed jets in p+p, d+Au and Au+Au collisions at $\sqrt{s_\mathrm{NN}} = 200~\mathrm{GeV}$ from the STAR experiment at RHIC. 
Measurement of initial state nuclear effects in d+Au collisions utilizing di-jet azimuthal correlations is presented together with similar measurement in p+p collisions.
Inclusive jet $\pt$ spectra and fragmentation functions in p+p and central Au+Au collisions are reported, with subsequent studies of jet nuclear modification factor, jet energy profile and modifications in the fragmentation function due to jet quenching.

}

\FullConference{European Physical Society Europhysics Conference on High Energy Physics\\
			 	   July 16-22, 2009\\
				   	 Krakow, Poland}

\usepackage{subfigure}
\newcommand{\pt}{p_\mathrm{T}}
\newcommand{\kt}{k_\mathrm{T}}
\newcommand{\et}{E_\mathrm{T}}
\newcommand{\gsim}{\,{\buildrel > \over {_\sim}}\,}
\newcommand{\gev}{\mathrm{GeV}}
\newcommand{\gevc}{\mathrm{GeV}/c}

\begin{document}

\section{Introduction}
\label{intro}

Jets are remnants of hard-scattered partons, which are the fundamental objects of pQCD. At RHIC, they can be used as a probe of the hot and dense matter created in heavy ion collisions. Baseline measurements in elementary collision systems can be used to isolate initial state effects from medium modification. In particular, measurement of the nuclear $\kt$ broadening~\cite{vitev}, the transverse momentum of a jet pair, in d+Au collisions is important.

Interaction and energy loss in the medium leads to jet quenching in heavy ion collisions. Until recently, it was studied indirectly using single particle spectra and di-hadron correlations~\cite{quenching}. These measurements are however limited in the sensitivity to probe partonic energy loss mechanisms due to well known biases~\cite{trenk}.  

Developments in theory (for example~\cite{fj,bgsub}) and experiment (STAR detector upgrades, increased RHIC luminosity) finally enabled full jet reconstruction in heavy ion collisions~\cite{initial}. Full jet reconstruction reduces the biases of indirect measurements and enables access to qualitatively new observables such as energy flow and fragmentation functions. We will present and discuss recently reported measurements of jet spectra~\cite{MP} and fragmentation functions~\cite{EB}.

\section{Jet reconstruction}
\label{jets}
The present analysis is based on $\sqrt{s_\mathrm{NN}} = 200~\gev$ data from the STAR experiment, recorded during 2006-2008. The Barrel Electromagnetic Calorimeter (BEMC) detector is used to measure the neutral component of jets, and the Time Projection Chamber (TPC) detector is used to measure the charged particle component of jets. In the case of a TPC track pointing to a BEMC tower, its momentum is subtracted from the tower energy to avoid double counting (electrons, MIP and possible hadron showers in the BEMC). To reduce possible BEMC backgrounds, the jet neutral energy fraction is required to be within $(0.1,0.9)$.

Recombination jet algorithms kt and anti-kt, part of the FastJet package~\cite{fj}, are used for jet reconstruction. To subtract the background, a method based on active jet areas~\cite{bgsub} is applied event-wise: $\pt^{Rec} = \pt^{Candidate} - \rho \cdot A$, with $\rho$ estimating the background density per event and $A$ being the jet active area. To reject fake jets (``jets'' found by jet finder unrelated to hard scattering), one has to either select sufficiently high jet $\pt$, or statistically subtract fake jet yield (see~\cite{MP,EB} for details).

\section{Nuclear $\kt$ broadening}
Run 8 p+p and d+Au data were used for this study, with BEMC high tower (HT) online trigger requirement (one tower with $\et > 4.3~\gev$). For d+Au system, the Beam Beam Counter detector in the Au nucleus fragmentation region was used to select the 20\% highest multiplicity events. An upper $\pt < 15~\gevc$ cut was applied to TPC tracks due to uncertainties in TPC tracking performance at high-$\pt$. 

Resolution parameter $R=0.5$ was used for jet finding and a cut $\pt > 0.5~\gevc$ applied for tracks and towers to reduce background. Due to the asymmetry of the colliding d+Au system, the background is asymmetric in $\eta$. This dependence was fit with a linear function in $\eta$ and included in the background subtraction procedure. To study detector effects, Pythia 6.410 and GEANT detector response simulations were used. Jet reconstruction was performed at MC level (PyMC) and at detector level (PyGe). To study effects of the d+Au background, PyGe events were added to minimum bias triggered d+Au events (PyBg). 

To select a clean di-jet sample two highest energy jets ($p_\mathrm{T,1} > p_\mathrm{T,2}$) in each event were used, with $p_\mathrm{T,2} > 10~\gevc$~\cite{myQM09}. Distributions of $k_\mathrm{T,raw} = p_\mathrm{T,1} \sin(\Delta\phi)$ were constructed for di-jets and Gaussian $\sigma_{k_\mathrm{T,raw}}$ was measured for the two jet algorithms and two ($10 - 20~\gevc$, $20 - 30~\gevc$) $p_\mathrm{T,2}$ bins. 

\begin{figure}[htb]
\begin{minipage}[h]{0.59\textwidth}  
\centering
\includegraphics[width=\textwidth]{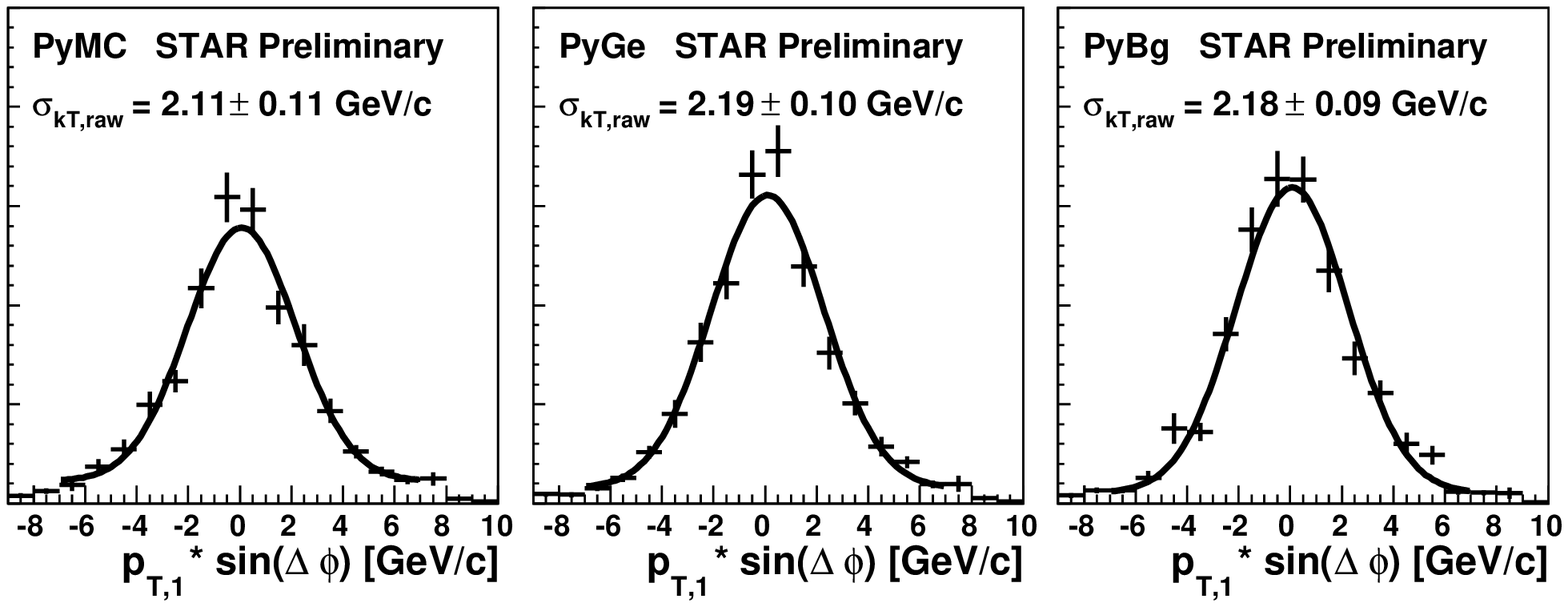}
\vspace{-0.9cm}
\caption{Distributions of $k_\mathrm{T,raw} = p_\mathrm{T,1} \sin(\Delta\phi)$ for simulation (anti-kt algorithm, $10 < p_\mathrm{T,2} < 20~\gevc$).}
\label{fig:ktsimu}
\end{minipage}
\hfill
\begin{minipage}[h]{0.39\textwidth}
\includegraphics[width=\textwidth]{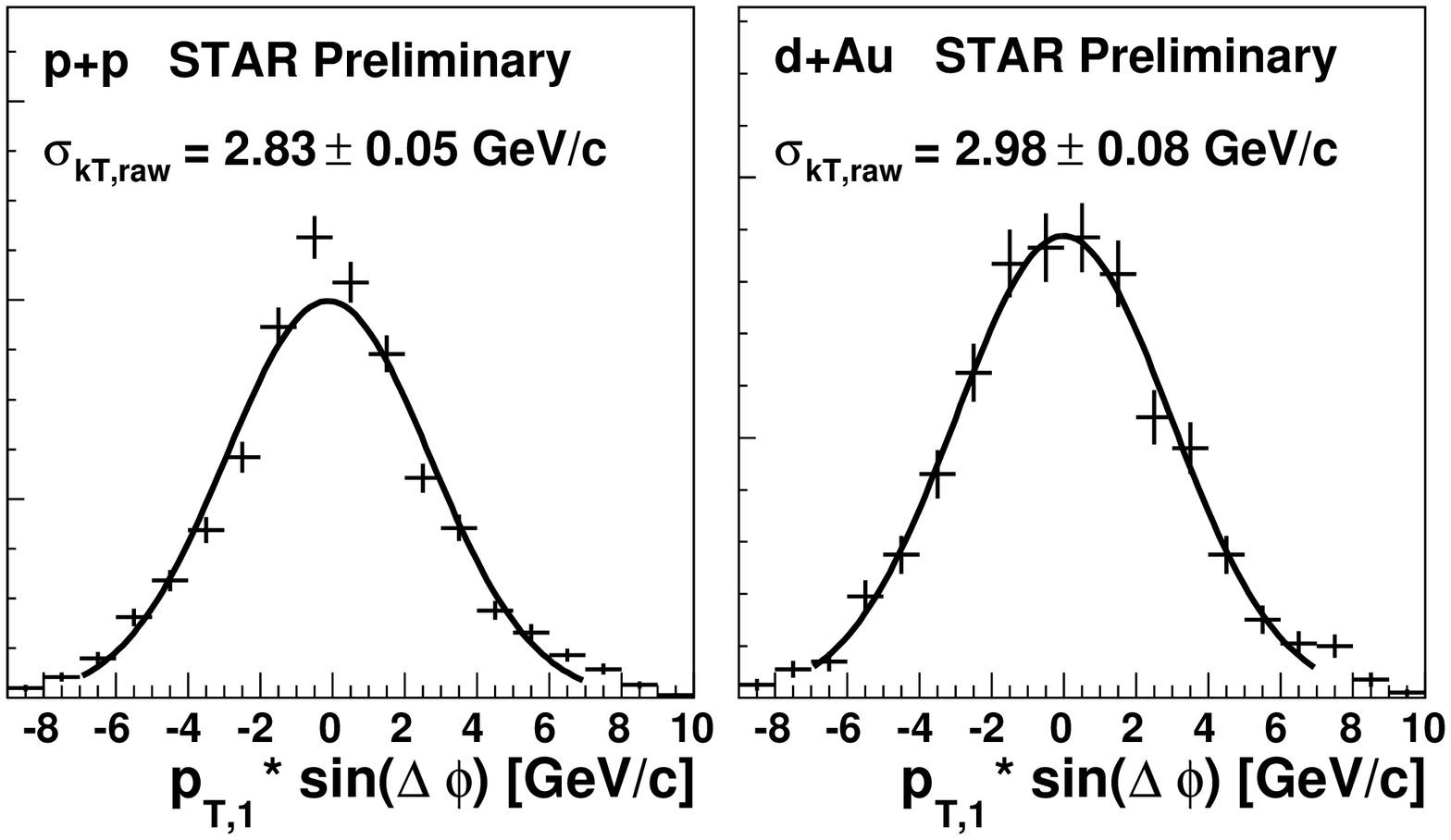}
\vspace{-0.9cm}
\caption{Distributions of $k_\mathrm{T,raw}$ for p+p and d+Au data, same cuts as in Figure~\protect \ref{fig:ktsimu}.}
\label{fig:ktdata}
\end{minipage}
\hfill
\end{figure}

Figure~\ref{fig:ktsimu} shows that the sigma widths are similar for PyMC, PyGe and PyBg distributions, due to the interplay between jet $\pt$ and $\Delta\phi$ resolutions. Figure~\ref{fig:ktdata} shows an example of the $k_\mathrm{T,raw}$ distributions for data. The extracted values are $\sigma_{k_\mathrm{T,raw}}^{p+p} = 2.8 \pm 0.1~\mathrm{(stat)}~\gevc$ and $\sigma_{k_\mathrm{T,raw}}^{d+Au} = 3.0 \pm 0.1~\mathrm{(stat)}~\gevc$. Possible nuclear $\kt$ broadening therefore seems rather small.

The systematic errors coming from a small dependence on the $|\Delta\phi - \pi|$ cut for back-to-back di-jet selection (varied between 0.5 and 1.0) and differences between the various selections ($p_\mathrm{T,2}$ range, jet algorithm) are estimated to be $0.2~\gevc$.
In addition, uncertainties in BEMC calibration and TPC performance in the high luminosity environment in run 8 are under study. Given the similarities between p+p and d+Au datasets, these are expected to be highly correlated.

\section{Inclusive jet spectra}
\label{sec_spectra}
Most central (0-10\%) minimum bias triggered Au+Au data from run 7 and jet patch triggered p+p data from run 6 were used for this study. Detail on trigger and correction for its bias can be found in~\cite{MP}. To minimize fragmentation biases, a minimal cut $\pt > 0.2~\gevc$ was used for tracks and towers. To study the jet energy profile, two resolution parameters ($R = 0.2$, $R = 0.4$) were used.

To be able to compare jet $\pt$ spectrum in Au+Au to p+p collisions, one has to correct for additional smearing due to background fluctuations. They are assumed to be Gaussian with width obtained by embedding Pythia jets into Au+Au events. Further corrections ($\pt$ independent tracking efficiency, jet energy resolution, unobserved neutral energy) were applied~\cite{MP}.

\begin{figure}[htb]
\begin{minipage}[h]{0.46\textwidth}  
\centering
\includegraphics[width=0.85\textwidth]{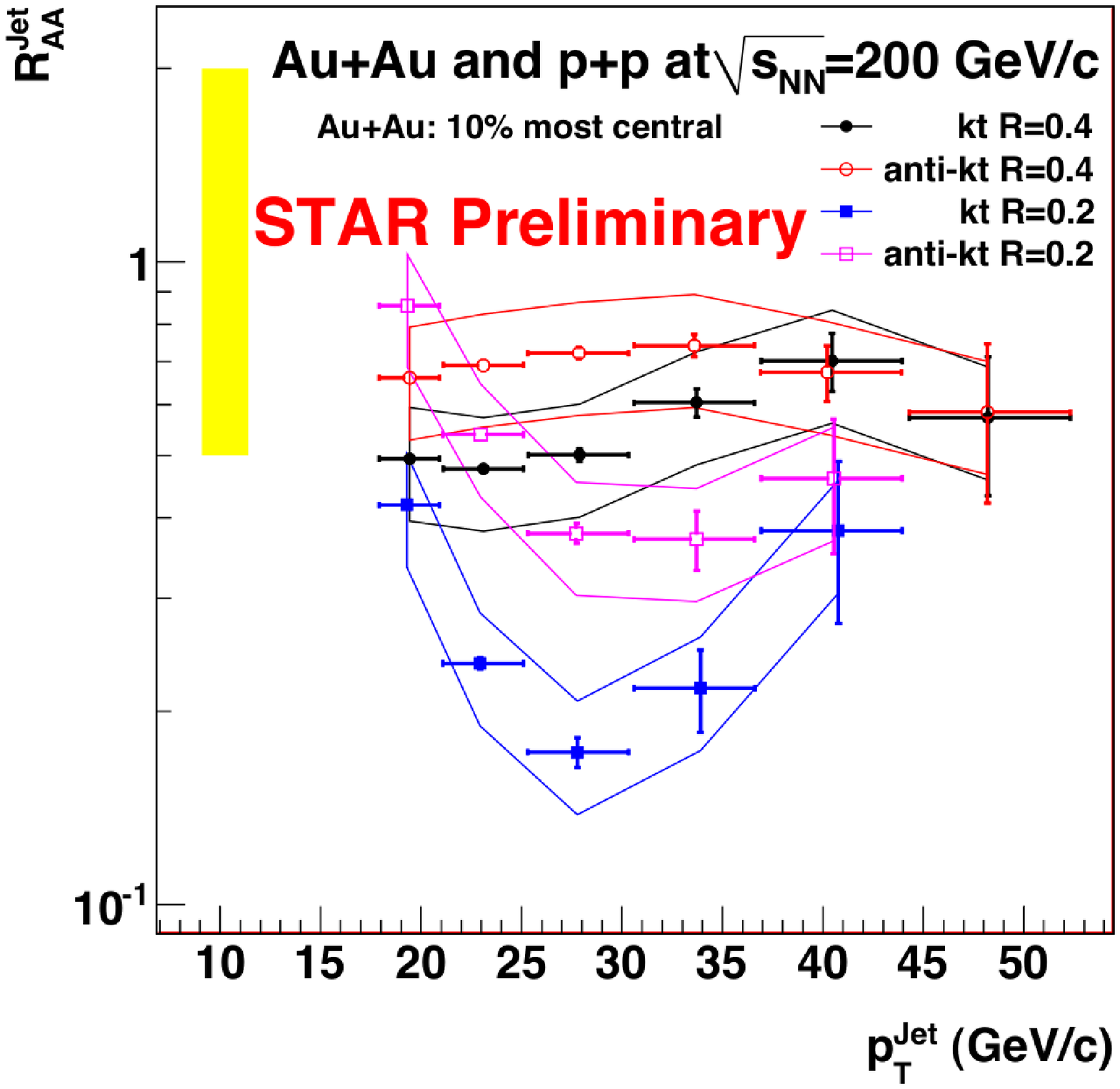}
\vspace{-0.6cm}
\caption{Jet $R_\mathrm{AA}$, yellow band shows jet energy scale uncertainty. Taken from~\cite{MP}.}
\label{fig:Raa}
\end{minipage}
\hfill
\begin{minipage}[h]{0.46\textwidth}
\includegraphics[width=0.85\textwidth]{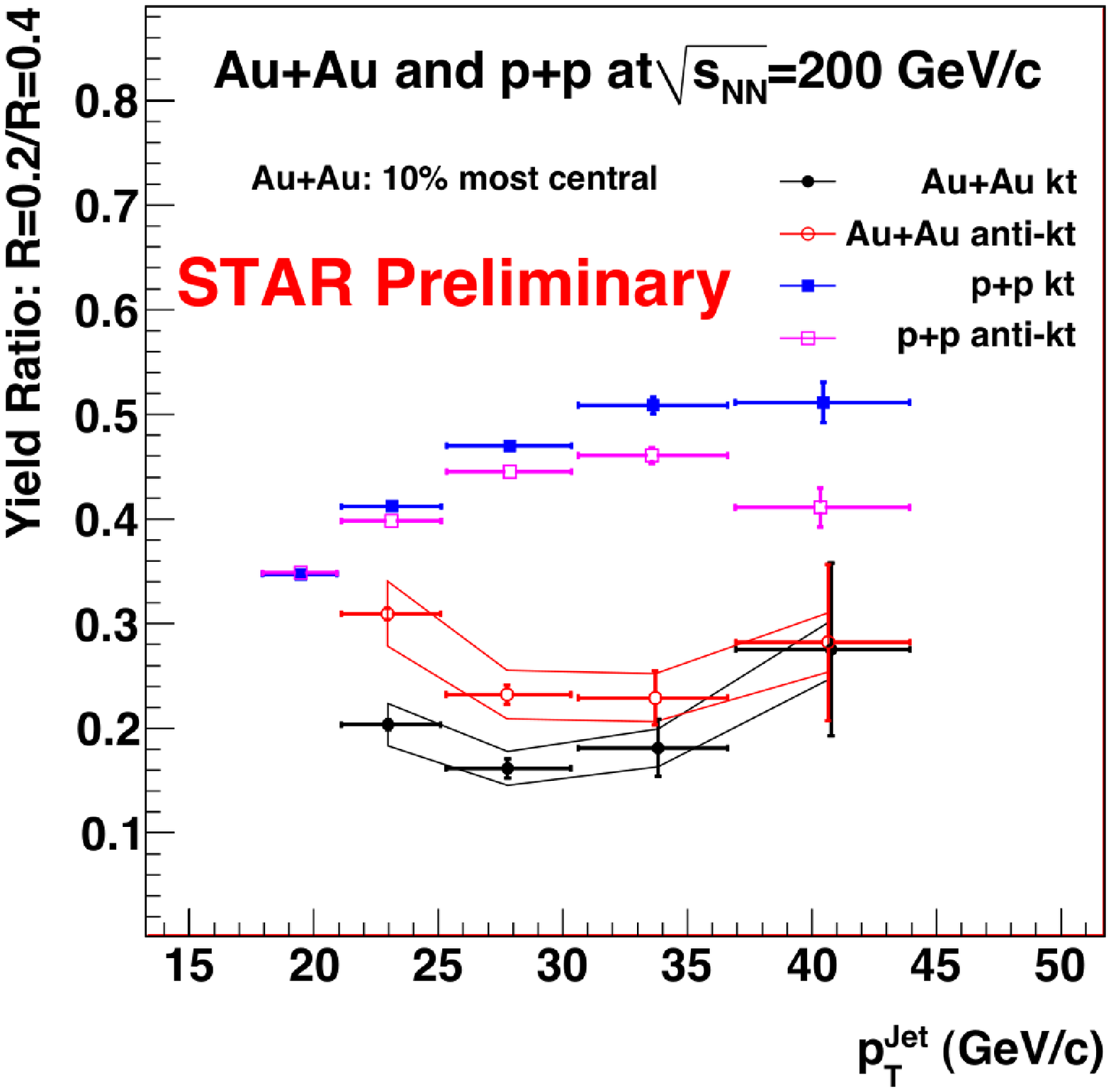}
\vspace{-0.6cm}
\caption{Ratio of jet yields for $R = 0.2, 0.4$ in p+p and Au+Au collisions. Taken from~\cite{MP}.} 
\label{fig:Rratio}
\end{minipage}
\hfill
\end{figure}

Figure~\ref{fig:Raa} shows $R_\mathrm{AA}^\mathrm{jet}$, the ratio of jet yield in Au+Au over the binary collision scaled jet yield in p+p. $R_\mathrm{AA}^\mathrm{jet}$ is expected to be close to unity for unbiased jet reconstruction (with possible small deviations due to initial state effects). 
For $R = 0.4$, $R_\mathrm{AA}^\mathrm{jet}$ is compatible with unity with large uncertainties, while for $R = 0.2$, the jets are significantly suppressed. 
The differences between kt and anti-kt algorithms are most likely due to different sensitivity to heavy ion background.
Figure~\ref{fig:Rratio} shows the ratio of jet $\pt$ spectra for $R = 0.2$ over that for $R = 0.4$ separately for p+p and Au+Au. The ratio is strongly suppressed in Au+Au with respect to p+p, indicating substantial broadening of jet structure in heavy ion collisions.

\section{Fragmentation function measurement}
Insight into jet structure is provided by studying the jet fragmentation function (FF). In-medium softening of the FF with respect to p+p reference should be observable in Au+Au for an unbiased jet population~\cite{softening}. Recoil jets on the away side of a BEMC HT triggered ($\et > 5.4~\gev$) jet are used for this study, to maximize the medium path length (and therefore possible FF modification). The online HT trigger condition is the same for p+p (run 6) and 0-20\% most central Au+Au (run 7) data. 

Anti-kt algorithm with a resolution parameter $R = 0.4$ was used. The jet $\pt$ is not corrected for instrumental effects yet and is marked $p_\mathrm{T, rec}$. Further information on corrections and unfolding can be found in~\cite{EB}. For trigger jets, a cut $p_\mathrm{T, rec} > 10~\gevc$ was applied to minimize fake jet contribution and $\pt > 2~\gevc$ was applied to tracks and towers to achieve similar jet energy scale in p+p and Au+Au. On the other hand, a minimal $\pt > 0.15~\gevc$ cut on tracks and towers was applied to recoil jets to minimize the biases. 
The recoil jet energy was determined using a resolution parameter $R = 0.4$, whereas charged hadrons in a larger cone ($R = 0.7$) around the jet axis were used to construct the FF. Background of the FF was subtracted event-wise using the charged hadron $\pt$ distribution in out of cone area.

\begin{figure}[htb]
\begin{minipage}[h]{0.45\textwidth}  
\centering
\includegraphics[width=\textwidth]{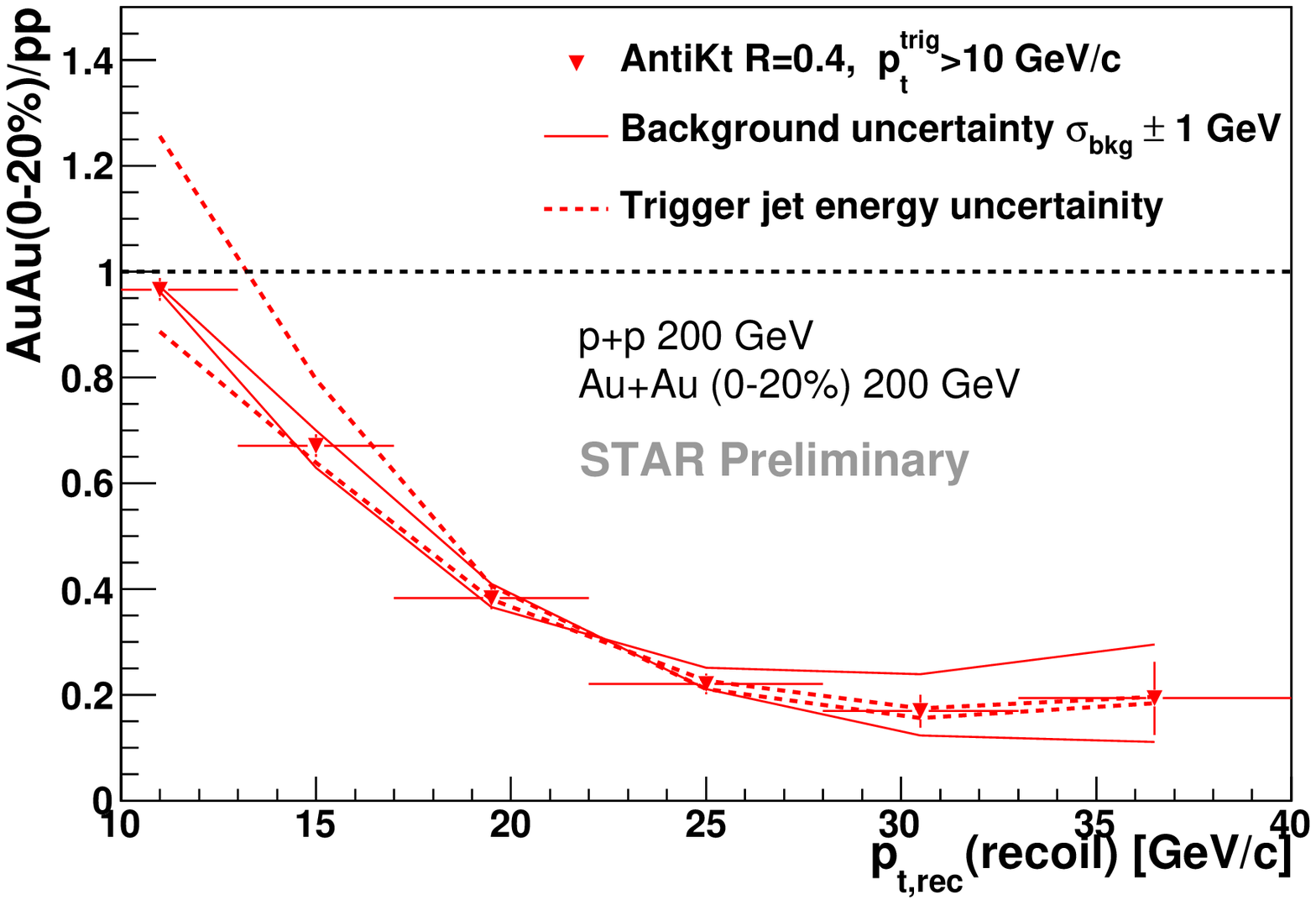}
\vspace{-0.8cm}
\caption{Ratio of recoil jet $\pt$ spectra in Au+Au to p+p. Taken from~\cite{EB}.}
\label{fig:dijet}
\end{minipage}
\hfill
\begin{minipage}[h]{0.45\textwidth}
\includegraphics[width=0.95\textwidth]{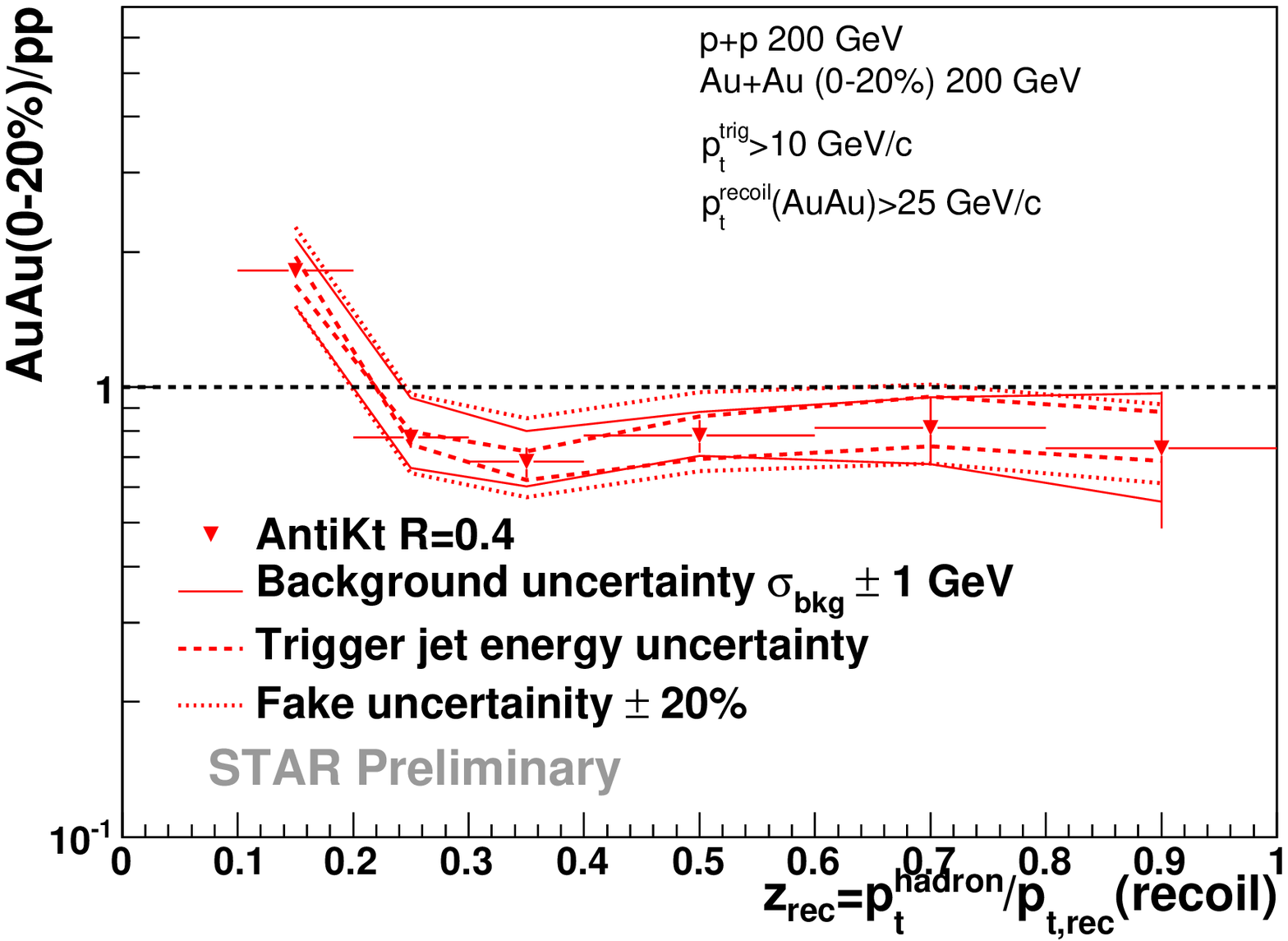}
\vspace{-0.3cm}
\caption{Ratio of $z_\mathrm{rec} = p_\mathrm{T}^\mathrm{hadron} / p_\mathrm{T, rec}^\mathrm{recoil}$ distributions in Au+Au to p+p. Taken from~\cite{EB}.}
\label{fig:FF}
\end{minipage}
\hfill
\end{figure}

Figure~\ref{fig:dijet} shows the ratio of Au+Au to p+p di-jet spectra, indicating a strong suppression of recoil jets in Au+Au. This is in contrast to the expected value of unity for unbiased jet reconstruction. Figure~\ref{fig:FF} shows the ratio of Au+Au to p+p recoil jet FF for $p_\mathrm{T, rec}^\mathrm{recoil} > 25~\gevc$. No strong modification of the fragmentation function for $z_\mathrm{rec} \gsim 0.2$ is observed. 

Significant suppression of recoil jets together with the absence of strong FF modification in Au+Au could be explained by broadening of the quenched recoil jets, causing an energy shift towards lower energies and artificial hardening of FF.
Alternatively, a part of recoil jet population may not be recovered due to strong broadening. In this case the measured FF in Au+Au may be that of surviving recoil jets, i.e. those with little modification (such as tangential emission or punch through jets)~\cite{EB}.

\section{Summary}
\label{summary}

Results from full jet reconstruction in $\sqrt{s_\mathrm{NN}} = 200~\gev$ p+p, d+Au and Au+Au collisions from the STAR experiment were presented. Study of jet $\kt$ broadening in p+p and d+Au collisions indicates that nuclear effects are rather small. Measurements of inclusive jet cross section as a function of resolution parameter and of recoil jet spectrum and fragmentation functions indicate strong broadening of the jet structure in heavy ion collisions.

\section*{Acknowledgement}
\label{acknowledgement}

This work was supported in part by GACR grant 202/07/0079 
and by grants LC07048 and LA09013 of the Ministry 
of Education of the Czech Republic.

\end{document}